\documentclass[fleqn,usenatbib,useAMS,linenumbers]{mnras}

   \makeatletter
\def\@fnsymbol#1{\ensuremath{\ifcase#1\or  \else\@ctrerr\fi}}
    \makeatother

\usepackage{subfigure}
\usepackage{fancyhdr}
\usepackage{booktabs}
\usepackage{graphicx}
\usepackage{threeparttable}
\usepackage{adjustbox}
\usepackage{csquotes}
\usepackage{arydshln}
\usepackage{lineno}
\usepackage{graphicx}		% Including figure files
\usepackage{amsmath}		% Advanced maths commands
\usepackage{amssymb}		% Extra maths symbols
\usepackage{multicol}        	% Multi-column entries in tables
\usepackage{bm}			% Bold maths symbols, including upright Greek
\usepackage{pdflscape}		% Landscape pages

\usepackage[T1]{fontenc}
\usepackage{ae,aecompl}

\usepackage{newtxtext,newtxmath}

\title[]{Variable white dwarfs in TMTS: Asteroseismological analysis of a ZZ Ceti star, TMTS J17184064+2524314}
 \author[Jincheng Guo et al.]{Jincheng Guo$^{1,\ast}$\thanks{$^{\ast}$\,E-mail: andrewbooksatnaoc@gmail.com (J. Guo), yonghuiyang1998@126.com (Y.Yang), wang\_xf@mail.tsinghua.edu.cn (X. Wang)}, 
Yanhui Chen$^{2,3,4}$,
Yonghui Yang$^{2,3,4,\ast}$,
 Xiaofeng Wang$^{5,1,\ast}$, 
 Jie Lin$^{6,7,5}$,
 Xiao-Yu Ma$^{8,9}$,\newauthor
 Gaobo Xi$^{5}$,
 Jun Mo$^{5}$,
 Alexei V. Filippenko$^{10}$,
 Thomas G. Brink$^{10}$,
 Weikai Zong$^{8,9}$, 
 Huahui Yan$^{11,12}$,\newauthor 
 Jingkun Zhao$^{12,13}$,
 Xiangyun Zeng$^{14}$,
 Zhihao Chen$^{5}$,
 Ali Esamdin$^{15}$,
 Fangzhou Guo$^{5}$,\newauthor 
 Abdusamatjan Iskandar$^{15}$, 
 Xiaojun Jiang$^{12,13}$, 
 Wenxiong Li$^{16}$,
 Cheng Liu$^{1}$,
 Jianrong Shi$^{12,13}$,\newauthor 
 Xuan Song$^{1}$,
 Letian Wang$^{15}$,
 Danfeng Xiang$^{5}$,
 Shengyu Yan$^{5}$,
 and Jicheng Zhang$^{9}$
 \\
$^{1}$Department of Scientific Research, Beijing Planetarium, Beijing 100044, China\\
$^{2}$Institute of Astrophysics, Chuxiong Normal University, Chuxiong 675000, China\\
$^{3}$International Centre of Supernovae (ICESUN), Yunnan Key Laboratory, Kunming 650216, China\\
$^{4}$Faculty of Science, Kunming University of Science and Technology, Kunming 650093, China\\
$^{5}$Physics Department and Tsinghua Center for Astrophysics, Tsinghua University, Beijing 100084, China\\
$^{6}$CAS Key laboratory for Research in Galaxies and Cosmology, Department of Astronomy,\\
University of Science and Technology of China, Hefei, 230026, People's Republic of China\\
$^{7}$School of Astronomy and Space Sciences, University of Science and Technology of China, Hefei, 230026, People's Republic of China\\
$^{8}$Institute for Frontiers in Astronomy and Astrophysics, Beijing Normal University, Beijing 102206, P. R. China\\
$^{9}$Department of Astronomy, Beijing Normal University, Beijing 100875, China\\ 
$^{10}$Department of Astronomy, University of California, Berkeley, CA 94720-3411, USA\\
$^{11}$Shandong Provincial Key Laboratory of Optical Astronomy and Solar-Terrestrial Environment,\\
School of Space Science and Physics, Shandong University, Weihai 264209, China\\
$^{12}$Key Laboratory of Optical Astronomy, National Astronomical Observatories, Chinese Academy of Sciences, Beijing 100101, China\\
$^{13}$School of Astronomy and Space Science, University of Chinese Academy of Sciences, Beijing 100049, China\\
$^{14}$Center for Astronomy and Space Sciences, China Three Gorges University, Yichang, 443000, People's Republic of China \\
$^{15}$Xinjiang Astronomical Observatory, Chinese Academy of Sciences, Urumqi 830011, China \\
$^{16}$The School of Physics and Astronomy, Tel Aviv University, Tel Aviv 69978, Israel\\
}

\begin{document}
 \date{}
 \pagerange{\pageref{firstpage}--\pageref{lastpage}} \pubyear{2023}
 \maketitle
 \label{firstpage}

\begin{abstract}
The Tsinghua University-Ma Huateng Telescope for Survey (TMTS)  has been constantly monitoring the northern sky since 2020 in search of rapidly variable stars. To find variable white dwarfs (WDs), the TMTS catalog is cross-matched with the WD catalog of Gaia EDR3, resulting in over 3000 light curves of WD candidates. The WD TMTS J17184064+2524314 (hereafter J1718) is the second ZZ~Ceti star discovered among these common sources. Based on the light curves from TMTS, follow-up photometric observations, and TESS, 10 periods and 3 combination periods are detected. A rotation period of $25.12\pm0.18$\,hr is derived, according to the identified rotational splitting. Our spectroscopic observation indicates that this WD belongs to DA type with $T_{\rm eff}=11,670\pm604$\,K, log\,$g=8.16\pm0.36$, $M = 0.70\pm0.23$\,M$_{\odot}$, and age=$0.51\pm0.34$\,Gyr. Based on core-parameterized asteroseismological model grids ($\geqslant$\,14 million), we derive a best-fit solution of $T_{\rm eff}=11,640\pm20$\,K, log\,$g=8.267\pm0.008$, and $M = 0.750\pm0.005$\,M$_{\odot}$ for J1718, consistent with the spectral fitting results. For this WD, the corresponding carbon and oxygen abundances in the core are 0.43 and 0.57, respectively. The distance derived from the intrinsic luminosity given by asteroseismology is $64\pm15$\,pc, in accord with the distance of $70.1\pm0.2$\,pc from Gaia DR3 within the uncertainties. 

\end{abstract}


 \section{Introduction}
Main-sequence stars with masses less than 10\,M$_{\odot}$ \citep{Woosley2015}, depending on metallicity, will end up as white dwarfs (WDs). WDs can provide information of great importance in various research areas \citep{Fontaine2008,Winget2008,Althaus2010,Saumon2022}. Because of their simple cooling mechanism, it is not difficult to derive relatively accurate ages of WDs. This makes them suitable for studies on the ages of Galactic populations \citep{Kilic2019,Guo2019}, which are important tools for studying their mass functions \citep{Holberg2016,Hollands2018} and luminosity functions \citep{Munn2017,Lam2019}. Accurate WD luminosity functions can be further used to probe the structure and evolution of the Galactic disk and open/globular clusters \citep{Bedin2009,Campos2016,Garca2016,Kilic2017,Guo2018}. Additionally, as potential gravitational-wave sources, ultrashort-period double WDs are likely to be detected by space gravitational-wave facilities like LISA \citep{Burdge2019}. Massive carbon-oxygen (CO) WDs in a binary system are the progenitors of Type Ia supernovae (SNe~Ia), which are standardizable candles for research in cosmology \citep[e.g.,][]{Maoz2014}. 

Studies of WDs have entered a new era promoted greatly by the increased numbers of stars \citep{Kleinman2013,Kepler2015,Kepler2019,Zhao2013,Gentile2015,Guo2015,Guo2022,Kong2021} from large spectroscopic surveys like Palomar-Green \citep{Green1986}, SDSS \citep{York2000}, and LAMOST \citep{Zhao2012}. But thanks to the Gaia mission \citep{Gaia2016}, studies of WDs now step into another new era. The number of WD candidates reached $\sim 359,000$ after the release of early Gaia DR3 \citep{Gentile2021}. These candidates are extremely useful for studies of photometric variability of WDs.

Asteroseismology can be used to study the internal structure of pulsating stars, including pulsating WDs (see a recent review by \cite{Corsico2019}). Ever since the discovery of HL Tau 76 \citep{Landolt1968}, the first pulsating hydrogen-rich WDs (DAV, or ZZ~Ceti star), almost 500 DAVs have been identified via photometric observations \citep{Corsico2022,Guo2023}, including those discovered with TESS \citep{Romero2022}. Many of these DAVs have been analysed with asteroseismology from different groups using different approaches and models. Two methods are generally used: (i) a set of fully evolutionary models, which are produced by tracking the complete evolution of the progenitor stars, from the zero-age main sequence to the WD stage \citep{Romero2012,Romero2013,Corsico2013,Geronimo2017,Geronimo2018}; and (ii) a set of static stellar models with parameterized chemical profiles, allowing a full exploration of the parameter space to locate an optimal seismic model, leading to good matches to the observed periods \citep{Bradley1998,Bradley2001,Castanheira2009,Bischoff2008,Bischoff2011,Bischoff2014,Giammichele2017a,Giammichele2017b,Guo2023}. Generally, the DAV instability strip ranges from $\sim 10,800$\,K to $\sim 12,000$\,K \citep{Gianninas2011,Van2012}. The atmospheres of these cool WDs are dominated by almost pure hydrogen, as a result of gravitational settling of other heavier elements. The hydrogen in the outer envelope recombines at $T_{\rm eff}\approx 12,000$\,K, resulting in a huge increase in envelope opacity. Accordingly, this restrains the flow of radiation and eventually causes $g$-mode oscillations. The excitation of $g$-mode oscillations in DA WDs at the blue edge of the instability strip could be due to the $\kappa$ mechanism, i.e. partial ionization of H \citep{Winget1982}. But a more physically plausible explanation is the convective driving mechanism \citep{Brickhill1991,Goldreich1999} when the outer convection zone deepens due to cooling. The oscillation-induced pulsation modes allow diagnosis of internal structures of these stars, based on the fact that they are quite sensitive to the stellar structure of DAVs. 

High-precision asteroseismology of WDs \citep{Ostensen2011,Hermes2017} began more than a decade ago \citep{Hermes2011,Greiss2016} with the Kepler mission \citep{Borucki2010} and the K2 mission \citep{Howell2014}. The sample has since been enriched by the TESS mission \citep{Ricker2015}, which monitors the entire sky \citep{Campante2016,Romero2022}. However, ground-based survey telescopes with short cadence are still essential for discovery and asteroseismological study of faint pulsating WDs.

In this work, we report the results of our asteroseismological analysis of the second DAV star (TMTS J17184064+2524314, dubbed as J1718) from TMTS after TMTS J23450729+5813146 \citep{Guo2023}. This star, with J2000 coordinates of $\alpha = 17^{\rm hr}18^{\rm m}40.64^{\rm s}$ and $\delta = +25^\circ24'31.4''$, was first reported as a WD candidate using SDSS DR7 photometry \citep{Girven2011}. The spectral energy distribution (SED) fitting suggests its effective temperature is 11,000\,K and surface gravity is 7.75. Similar results ($T_{\rm eff}=11,250$\,K, log\,$g$=7.50) were reported by \cite{Jimnez2018} by fitting the SED as well. Using accurate parameters from Gaia DR2, J1718 was found to be a thin-disk star by \cite{Torres2019}. The Gaia DR2 WD catalogue \citep{Gentile2019} includes the estimated parameters of J1718 as $T_{\rm eff}=11,595$\,K, log\,$g=8.05$, and $M=0.64$\,M$_{\odot}$ for hydrogen-rich atmosphere models, and $T_{\rm eff}=11,633$\,K, log\,$g=7.98$, and $M=0.57$\,M$_{\odot}$ for helium-rich atmosphere models. It was then studied in the Gaia 100\,pc WD sample \citep{Kilic2020}, where the estimation of parameters based on follow-up spectroscopy suggests that J1718 is a DA WD with $T_{\rm eff}=11,361\pm144$\,K, log\,$g=8.04\pm0.01$, and $M=0.63\pm0.01$\,M$_{\odot}$. 

J1718 was first identified as one of many new ZZ~Ceti WDs by \cite{Vincent2020}. According to two photometric follow-up observations lasting 5\,hr total, a main pulsation period of 731\,s was identified. Recently, variable WDs from Gaia and ZTF were studied by \cite{Guidry2021}. J1718 was observed by the McDonald 2.1\,m telescope, where three periods of 494.0\,s, 397.3\,s, and 531.9\,s were detected. Moreover, three (495.2\,s, 397.5\,s, 532.1\,s) and two (495.2\,s, 401.2\,s) periods were detected using data from the ZTF $g$ and $r$ bands, respectively.

This paper is organized as follows. Sect.\,2 presents the observational details of J1718. In Sect.\,3, frequency solution and asteroseismological modeling of J1718 are presented. Discussions and a summary are given in Sect.\,4.

\section{Observations}
\subsection{TMTS}

Located at Xinglong station in China, TMTS is a photometric survey with four 0.40\,m optical telescopes. With a total field of view of $\sim 18$\,deg$^{2}$ (4.5\,deg$^{2}$ $\times4$), it is equipped with four 4k $\times$ 4k pixel CMOS cameras, achieving a read-out time of $<1$\,s and high-speed photometry. Over the past three years, the main survey mode conducted on TMTS is the uninterrupted observations of LAMOST \citep{Cui2012} sky areas for the whole night with a cadence of 1\,min, which allows the TMTS to coordinate with the spectroscopic survey from the LAMOST observations. In order to fully exploit its potential, a customized filter covering 3300--9000\,\AA\ is used for the survey. The 3$\sigma$ detection limit of TMTS can reach 19.4\,mag for a 1\,min exposure. More information regarding the performance and research of TMTS can be found in \cite{Zhang2020}, \cite{Lin2022}, \cite{Lin2023a,Lin2023b}, \cite{Guo2023}, and \cite{Liu2023}.

Since 2020, the TMTS has produced uninterrupted whole-night light curves for nearly 20 million objects. The WD catalog generated from Gaia EDR3 \citep{Gentile2021} was cross-matched with the TMTS catalog within $3''$, yielding TMTS light curves for almost 3000 WD candidates. Based on visual inspections, J1718 was selected as a pulsating WD candidate for further study. It was observed by TMTS on two nights (see details in Table\,\ref{obslog}), with the amplitude of variation being 0.2--0.3\,mag, which is large for a ZZ~Ceti (similar to J2345).

\begin{table}
\footnotesize
\begin{center}
\caption{Observation logs of J1718.}
\label{obslog}
\begin{tabular}{lllll}
\\ \hline
Telescope & Instrument & UTC date & Exp. (s) \\   \hline
 TMTS     &  CMOS &    2020-05-09 & 60      \\
          &               &   2020-06-13  & 60      \\
  \\
 SNOVA  &  CCD     &  2022-07-20 &  50       \\
    (0.4\,m)          &          &   2022-08-17 & 60       \\
              &          &   2022-08-18 & 60       \\
\\
   TESS   &    600-1000 nm    & 2022-05-18 to  & 20/120   \\   
                 &          & 2022-06-13/Sector 52 &        \\
\\
Shane & Kast   & 2022-08-05 &   1800  \\
(3\,m)  &    spectrograph  &  &     \\ \hline

\end{tabular}
\end{center}
\end{table}

\begin{figure*}
    \centering
    \includegraphics[width=1.0\textwidth]{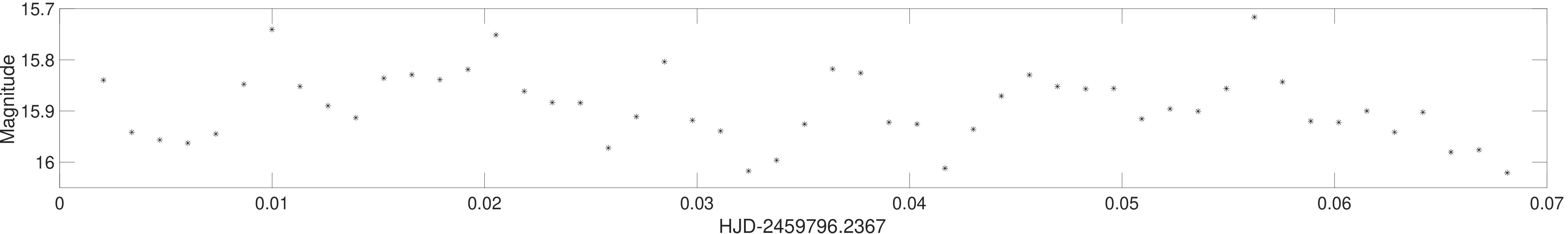}
    \caption{Part of the unfiltered light curve of SNOVA obtained in 2022.}
    \label{fig:lc}
\end{figure*}

\begin{figure}
    \centering
    \includegraphics[width=0.5\textwidth]{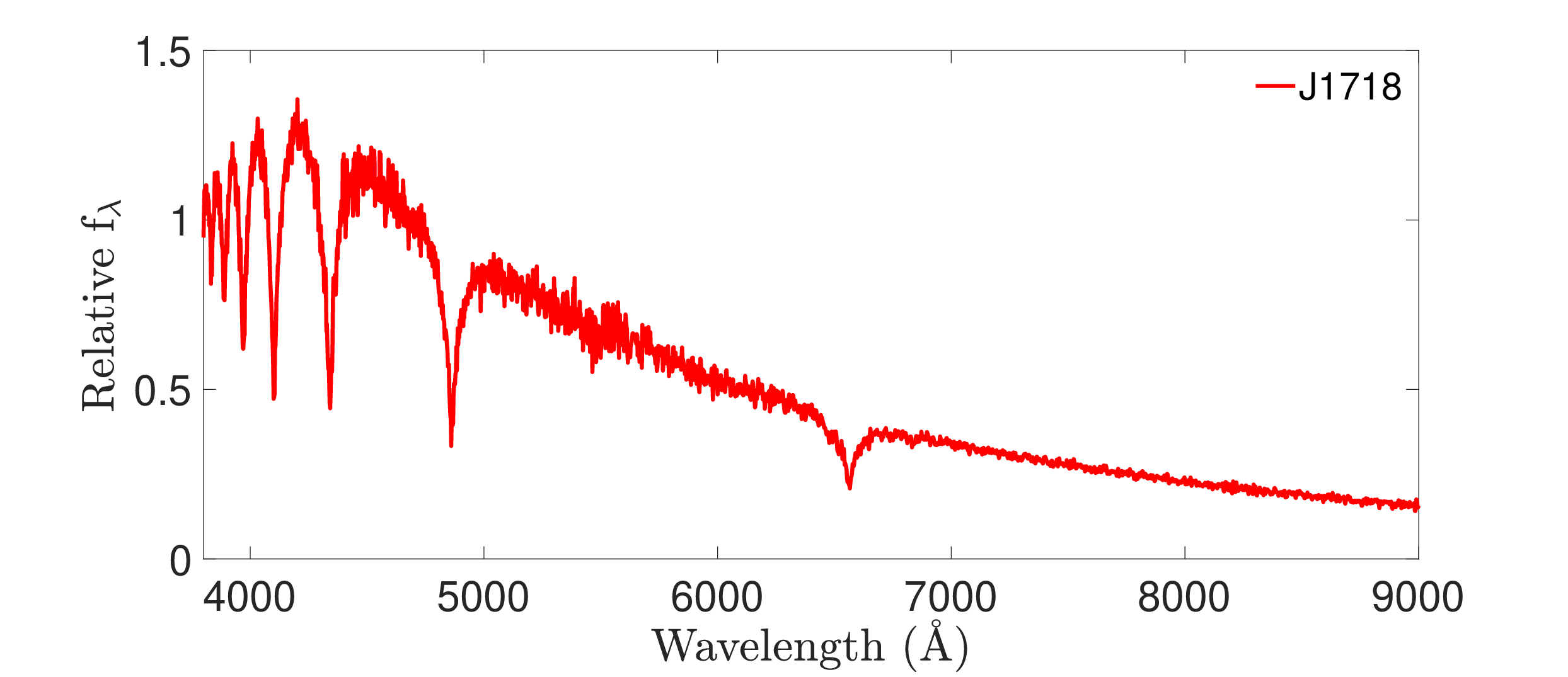}
     \includegraphics[width=0.4\textwidth]{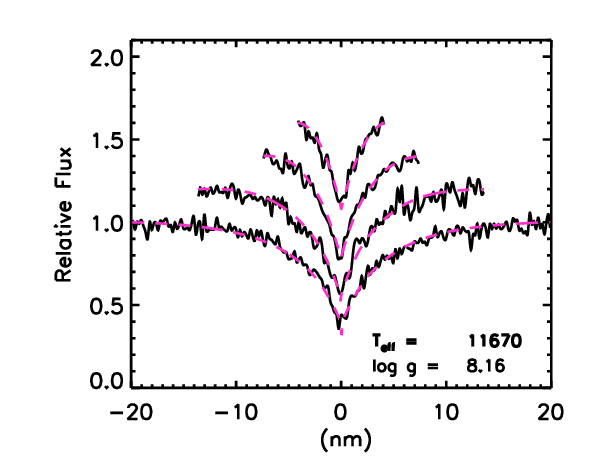}
    \caption{Upper panel: A spectrum taken by the Lick 3\,m Shane telescope (Kast spectrograph) on Aug. 05, 2022. 
    Lower panel: The spectral fits to the normalized H${\alpha}$, H${\beta}$, H${\gamma}$, and H${\delta}$ lines. The black lines represent the observed spectra, while the red dashed lines represent the best model fitting.}
    \label{spectrum}
\end{figure}

\subsection{Spectral observations}
A spectrum was taken on August 05, 2022 (UTC dates are used throughout this paper) with the Kast double spectrograph \citep{Miller1993} mounted on the 3\,m Shane telescope at Lick Observatory.
The data reduction with IRAF\footnote{IRAF: the Image Reduction and Analysis Facility is distributed by the National Optical Astronomy Observatory, which is operated by the Association of Universities for Research in Astronomy, Inc. (AURA) under cooperative agreement with the National Science Foundation (NSF).} follows standard procedure, including dark-current subtraction, bias subtraction, cosmic-ray removal, and one-dimensional (1D) spectral extraction. Following the spectral model fitting approach described by \cite{Guo2015,Guo2022}, the parameters of J1718 are derived to be $T_{\rm eff}=11,670\pm604$\,K, log\,$g=8.16\pm0.36$, $M=0.70\pm0.23$\,M$_{\odot}$, and cooling age $0.51\pm0.34$\,Gyr. The observed Balmer line profiles and the best-fit models are shown in Figure\,\ref{spectrum}. The derived $T_{\rm eff}$ and mass are especially helpful for providing an independent check with the results from asteroseismological analysis.

\subsection{Photometric follow-up observations}
Follow-up photometry of J1718 was obtained by using SNOVA, a 0.4\,m telescope located at Nanshan Observatory in China. The continuous  observations were obtained on three nights in white light (i.e., unfiltered) with an exposure time of $\sim 50$--60\,s. Standard image processing, such as bias correction, flat correction, and source extraction are performed with IRAF. Differential photometry was then applied to acquire the final light curves. Figure.\,\ref{fig:lc} displays part of the light curve taken by SNOVA. The light variation of J1718 is $\sim 0.2$--0.3\,mag, relatively large for a DAV. In addition to SNOVA, TESS observed J1718 in sector 52 from May 18 to June 13, 2022 (see Table\,\ref{obslog}). 

\section{Analysis}
\begin{figure}
    \centering
    \includegraphics[width=0.5\textwidth]{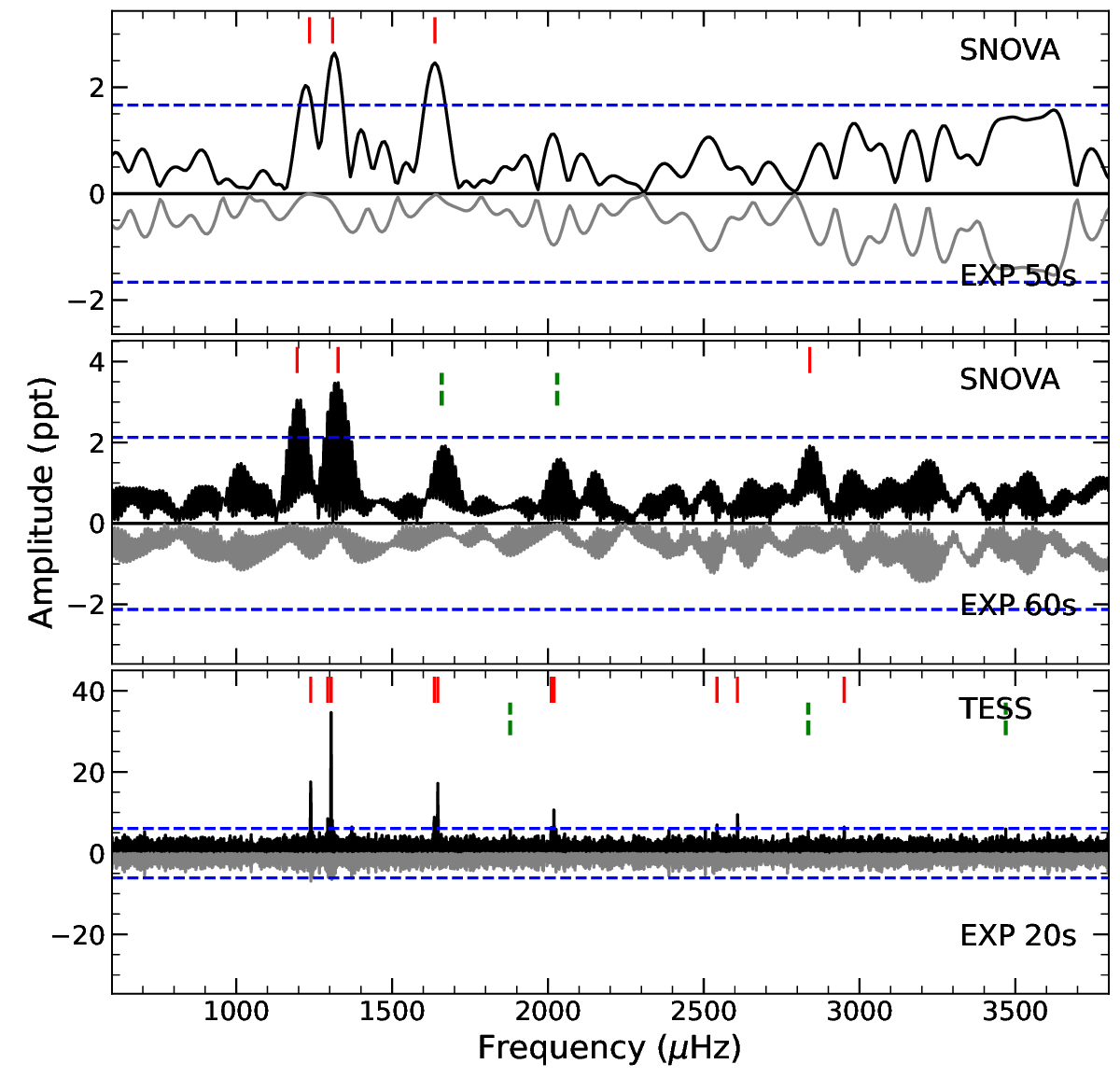}
    \caption{The Lomb-Scargle periodogram of J1718 for 50\,s and 60\,s observations by SNOVA and 20\,s exposures by TESS in 2022. The horizontal blue dashed lines are the 4.0$\sigma$ threshold. The red lines are identified frequencies above 4.0$\sigma$, while the green dashed lines are identified frequencies below but close to the 4.0$\sigma$ threshold. The upside-down grey curves represent the residuals of power spectra with significant frequencies being prewhitened.}
    \label{fig:FT}
\end{figure}

\begin{figure*}
    \centering
    \includegraphics[width=1\textwidth]{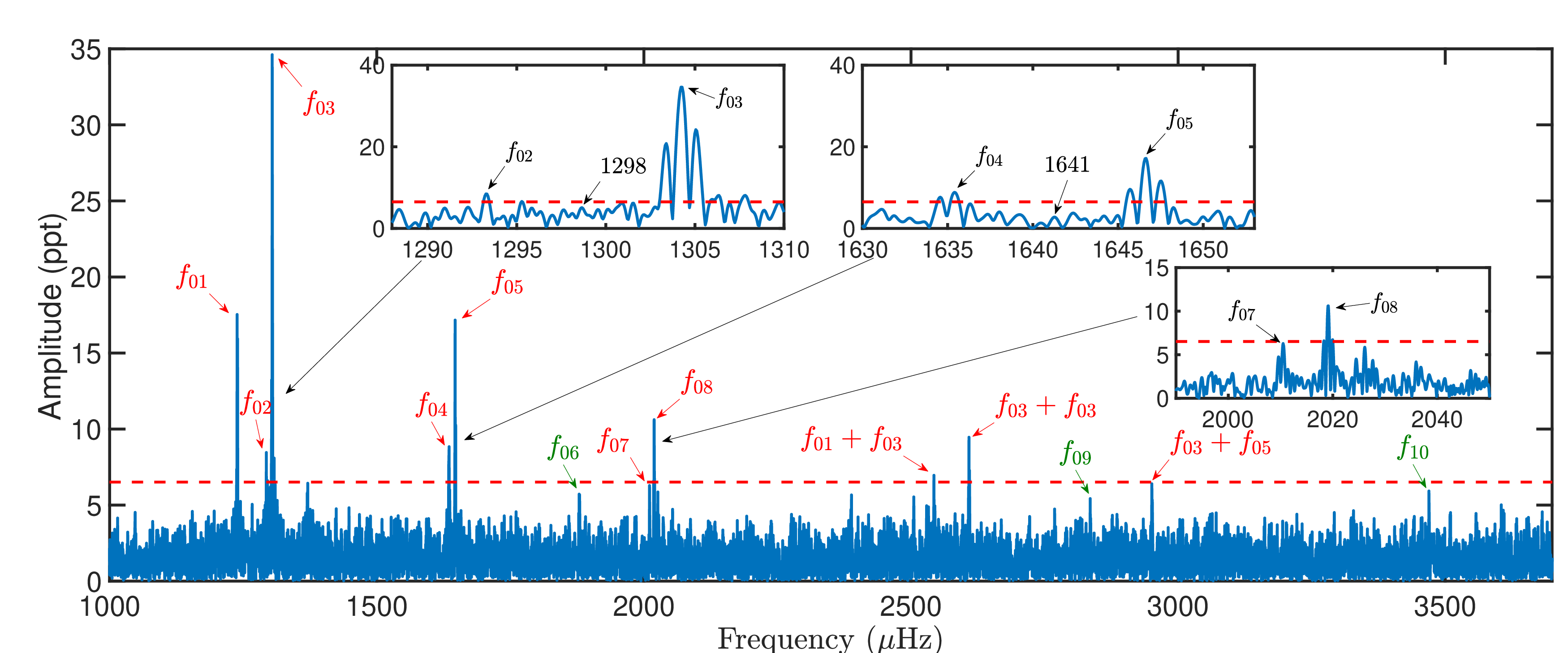}
    \caption{The zoom-in Lomb-Scargle periodogram of J1718 for 20\,s exposures by TESS in 2022. The horizontal red dashed line is the 4.0$\sigma$ threshold. The red text arrows are identified frequencies above 4.0$\sigma$, while the green text arrows are identified frequencies below but close to the 4.0$\sigma$ threshold. Details of three possible splitting frequency regions are shown in separate inset figures.}
    \label{fig:FT}
\end{figure*}

\begin{table*}
	\centering
	\caption{The frequency content detected from TESS 20\,s photometric observations.}
	\label{tab:1_table}
	\normalsize

    \begin{tabular}{cccccccc}
    	\hline
    	ID & Freq.           &  $\sigma {\rm Freq.}$ & $P$ & $\sigma P$ & Amplitude & $\sigma \rm{Amp}$ & SNR  \\
	    & ($\mu$Hz)   &   ($\mu$Hz)                  &  (s)        &  (s)              &  (ppt)         &    (ppt)                     &           \\ \hline

            \\
        $f_{01}$&1238.570&0.027  & 807.383&0.017&17.102&1.755&9.74\\
        $f_{02}$&1293.318& 0.057 & 773.205&0.034&8.068&1.766&4.57 \\ 
        $f_{03}$&1304.248&    0.013 & 766.725&0.008&34.231&1.751&19.55   \\ 
        $f_{04}$&1635.437&   0.045 & 611.457&0.017&9.593&1.635&5.87      \\ 
        $f_{05}$&1646.616&      0.024  & 607.306&0.009&17.569&1.633&10.76   \\ 
        $f_{06}^*$&1878.783& 0.066 & 532.259&0.019&5.814&1.465&3.97        \\ 
        $f_{07}$&2010.509&      0.059 &  497.386&0.015&6.564&1.486&4.42       \\ 
        $f_{08}$&2019.118&     0.037 &  495.266&0.009&10.531&1.475&7.14     \\ 
        $f_{09}^*$&2835.372&     0.070 & 352.687&0.009&5.645&1.522&3.71        \\ 
        $f_{10}^*$&3469.384&      0.066 & 288.236&0.005&5.980&1.511&3.96  \\  
        \\
        $f_{01}+f_{03}$&2542.866&      0.061  & 393.257&0.009&6.754&1.580&4.28     \\ 
        $f_{03}+f_{03}$&2608.428&      0.043 &  383.373&0.006&9.147&1.517&6.03      \\ 
        $f_{03}+f_{05}$&2950.704&         0.064 & 338.902&0.007&6.366&1.571&4.05        \\

        \hline

	\end{tabular}	
	\begin{tablenotes}
	    \item $^{*}$ Frequencies below the 4.0$\sigma$ detection limits.
	\end{tablenotes}
\end{table*}

\begin{table}
	\centering
	\caption{Period distribution in J1718. (Columns 2 and 4 are frequencies and their corresponding amplitudes, respectively. Column 3 is the frequency difference between split frequencies.)}
	\label{tab:2_table}
	\normalsize

    \begin{tabular}{ccccccc}
    	\hline
ID   &    	 Freq     & $\delta$ $f$     & A      & $P$  & $m$  & $l$     \\
      &   ($\mu$Hz)   & ($\mu$Hz) & (ppt)    &  (s)  &   &   \\ \hline
            \\
$f_{01}$ & 1238.570  &       & 17.102 & 807.383 & 0? & 1\,or\,2 \\
  \hdashline
$f_{02}$ &1293.318 &       & 8.068  & 773.205 & -1 & 1    \\
         &                     & 5.465 &        &         &    &      \\
         &1298.783 &       &        & 769.951  & 0  & 1    \\
         &                & 5.465 &        &         &    &      \\
$f_{03}$ &1304.248 &       & 34.231 & 766.725 & +1 & 1    \\
 \hdashline
$f_{04}$ &1635.437 &       & 9.593  & 611.457 & -1 & 1    \\
         &                      &5.590 &        &         &    &      \\
         &1641.027 &       &              &609.374 & 0   & 1    \\
         &                      &5.590 &        &         &    &      \\
$f_{05}$ &1646.616 &       & 17.569 & 607.306 & +1 & 1    \\
 \hdashline
$f_{06}$ &1878.783 &       & 5.814  & 532.259 & 0? & 1\,or\,2 \\
 \hdashline
$f_{07}$ &2010.509 &       & 6.564  & 497.386 & -1 & 2    \\
         &                     &8.609 &        &         &    &      \\
$f_{08}$ &2019.118 &       & 10.531 & 495.266 & 0  & 2    \\
 \hdashline
$f_{09}$ &2835.372 &       & 5.645  & 352.687 & 0? & 1\,or\,2 \\
$f_{10}$ &3469.384 &       & 5.98   & 288.236 & 0? & 1\,or\,2  \\  \hline

	\end{tabular}	
\end{table}

\begin{table}
	\centering
	\caption{Identification of the modes observed in J1718. The second column is the periods of our optimal model.}
	\label{tab:3_table}
	\normalsize

    \begin{tabular}{cccc}
    	\hline
$P_{\rm obs}$ & $P_{\rm model}$  & $l$ & $k$  \\ 
  (s)   & (s) &   &     \\ 
\hline
288.236  & 288.204 & 1 & 4  \\
352.687  & 353.635 & 2 & 12 \\
495.266  & 495.521 & 2 & 18 \\
532.259  & 532.222 & 1 & 10 \\
609.374   & 609.031 & 1 & 12 \\
769.951   & 769.841 & 1 & 16 \\
807.383  & 807.182 & 2 & 31 \\
\hline
	\end{tabular}	
\end{table}

\subsection{The amplitude spectrum and pulsation frequencies}

For the purpose of Fourier decomposition of the light curve, the Frequency Extraction for Light-curve exploitation (\texttt{FELIX}) software package was used \citep{Charpinet2010}, which is dedicated to Kepler and TESS photometry \citep{Zong2016,Charpinet2019}. This package includes standard prewhitening and nonlinear least-squares fitting techniques for Fourier transformation \citep{Deeming1975,Scargle1982}. The acceptance threshold for a significant frequency is set to be 4.0$\sigma$ of the local noise level \citep[e.g.,][]{1997A&A...328..544K}.

Figure\,\ref{fig:FT} shows the Lomb-Scargle periodograms of SNOVA (50\,s and 60\,s exposures) and TESS (20\,s exposure) observations; 5 and 13 frequencies were detected in the SNOVA and TESS observations, respectively, including 2 frequencies in SNOVA and 3 frequencies in TESS below the detection limit. Owing to the fact that five frequencies detected in SNOVA observations are all included in the TESS frequencies, all 13 frequencies detected in TESS data are used for further analysis. Table~\ref{tab:1_table} lists the frequencies with their attributes extracted with \texttt{FELIX}: the frequency ID (Column~1) ordered by their frequency value (Column~2), amplitude (Column~6) and uncertainty (Columns~3, 7), the corresponding period and uncertainty (Columns~4, 5), and the signal-to-noise ratio (SNR, Column~8). 

Among the 13 frequencies, 10 are significant and 3 are suspected (below 4.0$\sigma$, marked by the "$\ast$" symbol). According to our careful examination, three linear combination frequencies are identified within their uncertainties, which are listed in the last three rows in Table\,\ref{tab:1_table}.

\subsection{Period distribution}
Based on Table\,\ref{tab:1_table}, there are 10 independent signals left, after removing the linear combination frequencies. Since frequencies $f_{02}$ and $f_{03}$, $f_{04}$ and $f_{05}$, and $f_{07}$ and $f_{08}$ are very close, we can assume that they are components of two different triplets and one quintuplet, respectively. Reasonably, we can also assume that in the two triplets, two central modes are missing between $f_{02}$ and $f_{03}$, as well as between $f_{04}$ and $f_{05}$. Three modes are missing in the quintuplet, which contains $f_{07}$ and $f_{08}$. In order to carry out further analysis, the missing central modes of two triplets are averaged to be 1298.783\,$\mu$Hz and 1641.027\,$\mu$Hz. In this setting, they satisfy the relation that $[(5.465 \times 2) + (5.590 \times 2)]/4/8.609 = 0.642 \approx 0.6$. Meanwhile, there are equally spaced frequencies in the first two triplets. Therefore, we identify them as $l=1$ modes with the central frequencies of $m=0$ modes undetected. The third incomplete triplet is assumed to be the $l=2$ mode. The detailed mode identifications are listed in Table\,\ref{tab:2_table}.

\subsection{Rotational splitting}
As shown in Table\,\ref{tab:2_table}, there are two incomplete triplets identified as $l=1$ modes. They are interpreted as results of rotational splitting. An average value of frequency separation between the nearby two components of the triplets is derived as (5.465 + 5.465 + 5.590 + 5.590)/4 = 5.528\,$\mu$Hz. Considering the first order in the rotation angular frequency $\Omega$, the frequencies in the rotating case $\sigma_{k,l,m}$ ($m\neq 0$) are associated with the frequencies in the nonrotating case $\sigma_{k,l}$ ($m=0$) in the relation

\begin{equation}
\Omega=\frac{\left( \sigma_{k,l,m}-\sigma_{k,l} \right)}{m\times\left( 1-C_{k,l} \right)}\, ,
\label{eq:1}
\end{equation}
where $C_{k,l}=1/l(l+1)$ is in the asymptotic regime \citep{Brickhill1975}. By adopting this equation, the rotation period of J1718 is derived to be $25.12\pm0.18$\,hr.

\subsection{Input physics and model calculations}
In order to evolve grids for DAV models, version 16 of the White Dwarf Evolution Code \citep{Bischoff2018a} is applied. This code includes a new feature that enables modeling of the oxygen (O) profile in the core, which is parameterized by six parameters (i.e., $h1-h3$, $w1-w3$; see Fig.\,\ref{fig:oxygen}). This feature provides more accurate fitting results than previous versions of the code that did not include it. The equation of state and opacity tables from the Modules for Experiments in Stellar Astrophysics \cite[MESA; ][]{Paxton2018,Paxton2019} are adopted in the current version of \texttt{WDEC}. The standard mixing-length theory \citep{Bohm1971} and a mixing-length parameter of 0.6 \citep{Bergeron1995} are adopted in the analysis. 

The explored parameter spaces of evolved WD models are shown in Table\,\ref{ParaSpace}. To cover the initial ranges and crude steps in this table, a total of 14,486,688 WD models have been evolved. This is the first time that such a large number of evolved model grids are used for DAV model fitting. 
This applied model grid differs from those derived by considering the complete evolution of WDs in two ways. First, this model grid only considers the cooling phase of WDs, while fully evolutionary models take into account the complete evolution of the progenitor stars and the WD cooling phases. Second, fully evolutionary models are typically not used for model fitting of large samples, while the model adopted here constructs a large grids to fit the observed periods. Besides, this model includes parameterized central core as well. More description about the core parameters can be found in Figure\,1 of \cite{Bischoff2018a} and the user manual \citep{Bischoff2018b}. In the following fitting process, fine steps are used to optimize the best-fitting models found with crude steps.

\begin{table*}
\begin{center}
\caption{The explored parameter spaces of WDs evolved by \texttt{WDEC}.}
\label{ParaSpace}
\normalsize
\begin{tabular}{lllllllllll}
\hline
Parameters                             &Initial ranges     &Crude steps    &Fine steps     &Optimal values      &   2$^{nd}$ Optimal values \\
\hline
$M_{*}$/M$_{\odot}$                    &[0.500,0.850]      &0.010          &0.005          &0.750$\pm$0.005     &  0.800$\pm$0.005    \\
$T_{\rm eff}$ (K)                       &[10600,12600]      &250            &10             &11640$\pm$20              &  11090$\pm$10    \\
$-$log($M_{\rm env}/M_{\rm *}$)          &[1.50,2.00]        & 0.50          &0.01           &2.00$\pm$0.01     &  1.50$\pm$0.01  \\
                                       &[2.00,3.00]        & 1.00          &0.01           &                     \\
$-$log($M_{\rm He}/M_{\rm *}$)           &[2.00,5.00]        &1.00           &0.01           &3.00$\pm$0.01      &  2.00$\pm$0.01    \\
$-$log($M_{\rm H}/M_{\rm *}$)            & [4.00,10.00]       &1.00           &0.01           &6.00$\pm$0.01     &  8.00$\pm$0.01  \\
$X_{\rm He}$ in mixed C/He/H region    &[0.10,0.90]        &0.16           &0.01           &0.24$\pm$0.02    &  0.43$\pm$0.01 \\
\hline
$X_\mathrm{O}$ in the core             &                   &               &               &                  \\
\hline
$h1$                                     &[0.60,0.75]        &0.03           &0.01           &0.57$\pm$0.02       &    0.59$\pm$0.01\\
$h2$                                     &[0.65,0.71]        &0.03           &0.01           &0.66$\pm$0.02       &    0.61$\pm$0.01\\
$h3$                                     &0.85               &               &0.01           &0.83$\pm$0.03             &     0.90$\pm$0.01\\
$w1$                                     &[0.32,0.38]        &0.03           &0.01           &0.30$\pm$0.01       &    0.33$\pm$0.01\\
$w2$                                     &[0.42,0.48]        &0.03           &0.01           &0.49$\pm$0.02       &    0.42$\pm$0.01\\
$w3$                                     &0.09               &               &0.01           &0.13$\pm$0.02             &    0.13$\pm$0.01\\
\hline
\end{tabular}
\end{center}
\begin{tablenotes}
\item $M_{*}$ denotes the stellar mass, while $M_{\rm env}$ is the envelope mass. $M_{\rm H}$ is the mass of the hydrogen atmosphere, and $M_{\rm He}$ is the mass of the He layer. $X_{He}$ is the He abundance in the mixed C/He/H region, and $X_{O}$ is the O abundance in the core. The parameter $h1$ refers to the O abundance in the core center, while $w1$ is the mass fraction of $X_{O} = h1$. Parameters $h2$ and $h3$ are O abundance of two knee points on the declining O profile; $w2$ and $w3$ refer to the masses of the gradient regions of the O profile. 

\end{tablenotes}

\end{table*}

\begin{figure}
    \centering
    \includegraphics[width=0.5\textwidth]{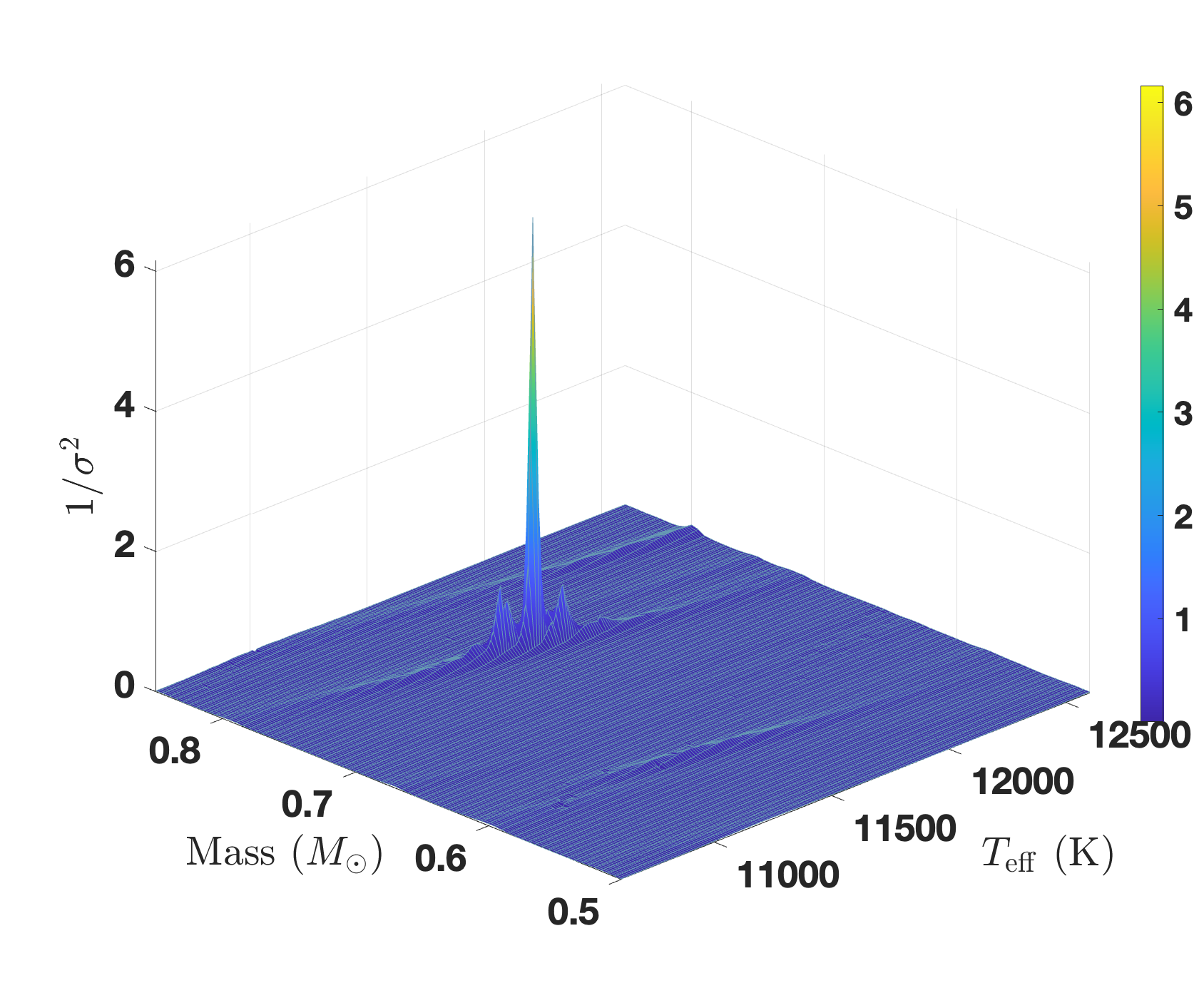}
    \includegraphics[width=0.5\textwidth]{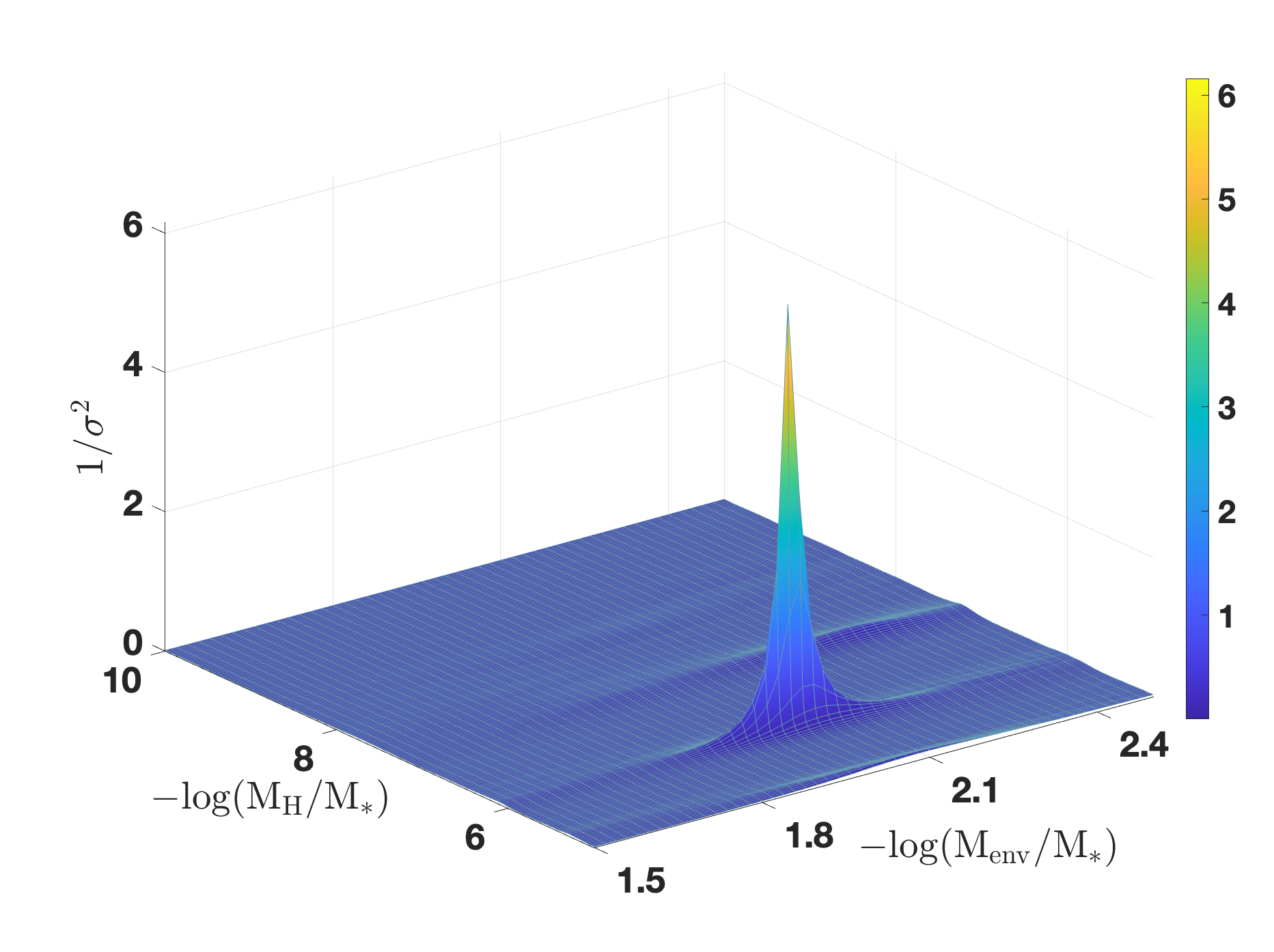}
    \caption{Upper panel: a color 3D graph of the fitting error $\sigma_{\rm RMS}$ as a function of stellar mass and $T_{\rm eff}$. A total of 14,271 DAV stellar models are used in the fit. Lower panel: a color 3D graph of the fitting error $\sigma_{\rm RMS}$ as a function of H-layer mass and envelope mass. A total of 12,801 DAV stellar models are used in the fit. }
    \label{fig:teffmass_envhe}
\end{figure}

\subsection{The model fittings}
\label{sec:fitting}

To evaluate the fitting results, the root-mean-square (RMS) residual $\sigma_{\rm RMS}$ is used,
\begin{equation}
\sigma_{\rm RMS}=\sqrt{\frac{1}{n}\sum_{n}^{}(P_{\rm obs}-P_{\rm mol})^{2}}\, ,
\label{eq:1}
\end{equation}
where $n$ represents the number of the fit modes (i.e., 7 for J1718). The observed and model periods are denoted as $P_{\rm obs}$ and $P_{\rm mod}$, respectively. During the model fittings, only the first 3 digits after the decimal point of the observed periods are used. Because the periods 773.205\,s and 766.725\,s, 611.457\,s and 607.306\,s, and 497.386\,s and 495.266\,s are very close, we assume that they are components of two rotational triplets and one quintuplet, respectively. Therefore, we use their central modes of 769.951\,s, 609.374\,s, and 495.266\,s, respectively. Together with 4 other periods, they are listed in the first column of Table\,\ref{tab:3_table} as the observed periods. 

In the initial fitting, the combinations of parameter space ranges and crude steps are used to build the DAV models. This includes 14,486,688 models, which are adopted to fit the observed 7 modes. The crude steps are chosen by considering the computational power for models. It takes about 7\,s to evolve one model using a 128-core/256-threads server with a clock speed of 2.45\,GHz. After finding the models with the minimum $\sigma_{\rm RMS}$, the parameter steps of the models are narrowed down to fine steps around those models. For the fine steps, there are 3 to 5 grid points for each parameter. The fine steps of mass are chosen to be 0.005 because this is the step size adopted in the WDEC. The fine steps of other parameters are chosen to be much smaller than their crude steps. There are 6 global parameters ($M_{*}$/M$_{\odot}$, $T_{\rm eff}$, $-$log($M_{\rm env}/M_{\rm *}$), $-$log($M_{\rm He}/M_{\rm *}$), $-$log($M_{\rm H}/M_{\rm *}$), and $X_{\rm He}$ in mixed C/He/H region) and 6 $X_{O}$ parameters ($h1-h3$, $w1-w3$) in Table\,\ref{ParaSpace}.

In order to find the minimum $\sigma_{\rm RMS}$, the $X_{O}$ parameters are fixed at first. Then by using fine steps, a grid of global parameters is generated to fit. Next, the global parameters are fixed to previously found values, and using fine steps, a grid of $X_{O}$ parameters is generated to fit. This process is repeated a dozen times, and the chosen fitting model converges to an optimal one.

In the actual fitting process, the optimal model using crude step grids is found to be $T_{\rm eff}=11,850$\,K and $M=0.85$\,M$_{\odot}$ with $\sigma_{\rm RMS}=1.06$, while the second optimal model is obtained to be $T_{\rm eff}=11,600$\,K and $M=0.75$\,M$_{\odot}$ with $\sigma_{\rm RMS}=1.21$. Owing to the fact that the mass of the second optimal model is more consistent with our spectral-fitting mass (i.e., 0.70\,M$_{\odot}$), we chose to find the final optimal model using fine steps based on the second optimal model. Then the final optimal model using fine steps is obtained when $\sigma_{\rm RMS} = 0.4$\,s, with the corresponding periods and other parameters listed in the second column of Table\,\ref{tab:3_table} and fifth column of Table\,\ref{ParaSpace}, respectively. The uncertainty of each optimal value is estimated by the full width at half height of the reciprocal of $n\sigma_{\rm RMS}$. According to the derived optimal values, J1718 is a relatively massive DAV with a slightly thick H atmosphere.

\begin{figure*}
    \centering
    \includegraphics[width=0.8\textwidth]{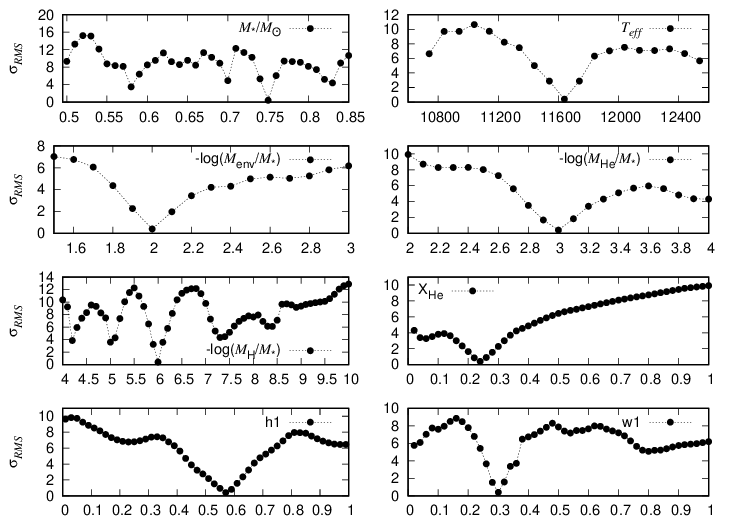}
    \caption{Sensitivity plots of 8 parameters of the optimal model. Subplots are 8 parameter values versus the residuals of their RMS, $\sigma_{\rm RMS}$. }
    \label{fig:dependence}
\end{figure*}

\begin{figure}
    \centering
    \includegraphics[width=0.5\textwidth]{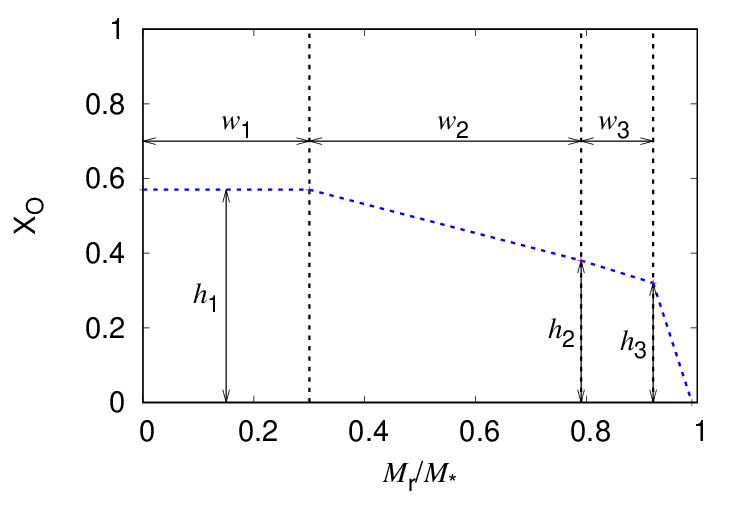}
    \caption{Chemical profile of oxygen as a function of the mass coordinate. Blue dashed line is the oxygen abundance profile in the core described by six parameters.}
    \label{fig:oxygen}
\end{figure}

\begin{figure}
    \centering
    \includegraphics[width=0.5\textwidth]{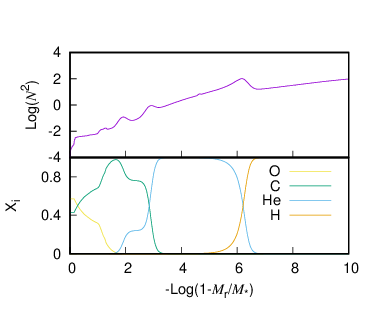}
    \caption{The core composition profiles and corresponding Brunt-V{\"a}is{\"a}l{\"a} frequency of the optimal model as functions of the mass coordinate logarithm. Upper panel: the logarithm of the square of buoyancy frequency as a function of stellar mass. Several bumps are present due to the existence of element transition gradient zones. Lower panel: the abundance of O, C, He, and H, plotted in yellow, green, blue, and brown lines, respectively. }
    \label{fig:2cxln2}
\end{figure}

The upper panel of Fig.\,\ref{fig:teffmass_envhe} shows a three dimensional (3D) color graph of the fitting error $\sigma_{\rm RMS}$ as a function of stellar mass and $T_{\rm eff}$. For the purpose of a better visual effect, the $z$-axis and colour represent the inverse of the fitting error $\sigma_{\rm RMS}$ (as defined in Eq.\ref{eq:1}) squared. The grid consists of stellar mass in the range of 0.500\,M$_{\odot}$ to 0.850\,M$_{\odot}$ with a step of 0.005\,M$_{\odot}$ and $T_{\rm eff}$ in the range of 10,600\,K to 12,600\,K with a step of 10\,K, which contains 14,271 DAV models. All other parameters are fixed to their optimal values in Table \ref{ParaSpace}. The lower panel of Fig.\,\ref{fig:teffmass_envhe} shows $\sigma_{\rm RMS}$ as a function of hydrogen layer mass and envelope mass, which involves 12,801 DAV models. The logarithm of the envelope mass ranges from $-1.50$ to $-2.40$ with a step of 0.02, and the logarithm of H layer mass ranges from $-5.00$ to $-10.00$ with a step of 0.02. The highest points in both figures correspond to values of our optimal $T_{\rm eff}$, mass, H-layer mass, and envelope mass.

Fig.\,\ref{fig:dependence} displays the sensitivity plots (fitting error $\sigma_{\rm RMS}$) of eight parameters of the optimal model, including the stellar mass, $T_{\rm eff}$, envelope mass, helium layer mass, hydrogen atmosphere mass, helium abundance in the mixed C/He/H region, and $h1$ and $w1$ for the O profile. Note that the sensitivity of each parameter is derived when all other parameters are fixed to the optimal values. The results with the lowest $\sigma_{\rm RMS}$ indicate the optimal values. In Table\,\ref{ParaSpace}, the mass of the H shell is in the range of $10^{-10}\,M_{\ast}$ to $10^{-4}\,M_{\ast}$. One should note that after an iteration, the fitting process may explore a broader range of values according to the pre-optimal solution. This is because we want to find the optimal values for the parameters, even if they are outside of the initial ranges listed in Table\,\ref{ParaSpace}.

Fig.\,\ref{fig:oxygen} is plotted according to six parameters, listed in Table\,\ref{ParaSpace}, that describe oxygen abundance in the core. The blue dashed line represents the oxygen profile as a function of the mass coordinate. The definitions of the six parameters are demonstrated in the figure. The core composition profiles and corresponding Brunt-V{\"a}is{\"a}l{\"a} frequency for the optimal model are shown in Fig.\,\ref{fig:2cxln2}. The bumps in the profile are caused by composition gradient in different zones. It is important to note that this kind of chemical profile is in contrast to those derived from the fully evolutionary computations. More detailed information regarding the existence of a C-buffer and several features in the chemical profile has been discussed extensively by \citet{Geronimo2019} for a DBV star. 

To check the self-consistency of the best-fit model, we can use the distance derived from asteroseismology analysis to perform an independent check. The luminosity of J1718 is log($L$/L$\odot$) $= -2.736\pm0.183$. Using the correction $M_{\rm bol} = M_{\rm bol,\odot} -2.5\times log(L/{\rm L}_{\odot})$ and assuming the bolometric magnitude of the Sun $M_{\rm bol,\odot} = 4.74$ \citep{Cox2000}, the bolometric magnitude of the optimal model is $M_{\rm bol}=11.58$\,mag. The $V$-band bolometric correction (BC) is adopted for J1718, as the Gaia $G$-band BC is not available in the DR3. As listed in Table\,1 of \cite{Bergeron1995a}, the $V$-band BC for a DA star model with $T_{\rm eff}= 11,000$\,K and 12,000\,K is $-$0.441\,mag and $-$0.611\,mag, respectively. Therefore, for J1718 with $T_{\rm eff}= 11,640$\,K, the linear interpolation of BC ($V$) is $-$0.550\,mag. Thus, the absolute $V$-band magnitude is derived to be $M_{V}=M_{\rm bol}-BC(V)=11.58-(-0.550)=12.13$\,mag. Given $m_{V}=16.152$\,mag \citep{Qi2015}, the distance inferred from the optimal model is $64\pm15$\,pc by using the equation $d=10^{(m_{V}-M_{V})/5+1}$. Within the quoted uncertainty, this result is consistent with the parallax-inversed distance of $70.1\pm0.2$\,pc from the Gaia DR3. 
The consistency of the asteroseismology analysis with that derived from the spectral model fitting indicates that our asteroseismological results are self-consistent.

\subsection{Alternative solutions}

In addition to the adopted solution, two more period interpretations are explored as well. In one scenario, we consider that top 10 frequencies listed in Table\,\ref{tab:1_table} are all independent frequencies without splitting. After repeating the same period-fit process mentioned above are applied, the final optimal model is obtained with fitting error $\sigma_{\rm RMS}$=1.26\,s. However, the parameters of $T_{\rm eff}$=12\,530\,K and mass=0.89\,M$_{\odot}$ differ a lot from parameters derived from spectral fitting results of $T_{\rm eff}$=11\,670\,K and mass=0.70\,M$_{\odot}$. Moreover, the derived asteroseismological distance is 59\,pc, which is 19\% less than Gaia distance. And there is only one period, 496.687\,s, exists in the optimal model in the range of 400-500\,s, while two periods of 495.266\,s and 497.386\,s are observed. Based on all the evidences, this scenario is ruled out.

In the second scenario, we assume that there are 7 independent frequencies, which are listed in Table\,\ref{tab:alter_table}. This time we consider that $f_{03}$, $f_{05}$ and $f_{08}$ are central frequencies of three different triplets, and for each triplet, one frequency is not detected in the TESS observation. After repeating the period-fit analysis, the optimal model for this scenario is found with $\sigma_{\rm RMS}$=0.09\,s. The corresponding parameters are listed in the last column of Table\,\ref{ParaSpace}. The parameters of optimal model are consistent with spectral fitting results within the uncertainties. But the derived asteroseismological distance of this solution is 58\,pc, which is 21\% less than Gaia distance of 70\,pc. Furthermore, among the 7 observed independent frequencies, there are four $l=2$ modes more than three $l=1$ modes in the optimal solution. Usually, $l=1$ mode should be easier to be detected than $l=2$ mode in an observation. So among the observed frequencies, it should be normal to find more $l=1$ modes than $l=2$ modes. Therefore, we still prefer the period interpretation in Table\,\ref{tab:2_table} and adopt the optimal values listed in the fifth column of Table\,\ref{ParaSpace} as the final optimal solution.

\begin{table}
	\centering
	\caption{Alternative identification of the modes observed in J1718. The second column is the periods of 2$^{nd}$ optimal model.}
	\label{tab:alter_table}
	\normalsize
	
    \begin{tabular}{cccc}
    	\hline
$P_{\rm obs}$ & $P_{\rm model}$  & $l$ & $k$  \\ 
  (s)   & (s) &   &     \\ 
\hline
288.236  & 288.161 & 2 & 9  \\
352.687  & 352.775 & 2 & 12 \\
495.266  & 495.144 & 1 & 9 \\
532.259  & 532.162 & 2 & 19 \\
607.306   & 607.334 & 1 & 12 \\
766.725   & 766.692 & 1 & 16 \\
807.383   & 807.268 & 2 & 30 \\
\hline
	\end{tabular}	
\end{table}

\section{Discussion and Summary}

Although a few pulsating periods of the ZZ~Ceti star J1718 were previously detected, we performed a thorough asterosesimological analysis for the first time with more extensive data, deriving accurate parameters for this pulsating white dwarf. 

Based on data from the TMTS and SNOVA, multiple pulsating periods were identified after prewhitening. However, since J1718 was observed by TESS in 2022, whose pulsating periods covered all TMTS and SNOVA periods, we only focus on the TESS-detected periods. There are 13 frequencies detected in TESS 20\,s data after prewhitening. Among those, 3 frequencies are linear combination frequencies. The remaining ten frequencies are divided into two incomplete triplets, one incomplete quintuplet, and four independent frequencies. The rotational period of $25.12\pm0.18$\,hr is derived using two triplets listed in Table\,\ref{tab:2_table}. 

The identified modes are used to perform the asteroseismological analysis. By adopting version 16 of \texttt{WDEC}, more than 14 million WD models were evolved for the parameter spaces listed in Table \ref{ParaSpace}. To find the optimal model, parameter steps and spaces are slowly narrowed down repeatedly, in accordance to the fitting result. Finally, the optimal model is found and its parameter values are listed in Table\,\ref{ParaSpace}. From the asteroseismological result, it can be learned that J1718 is a relatively massive pulsating WD with a slightly thick H atmosphere. The final results are consistent with the parameters derived from spectral fitting of our follow-up spectrum. In addition, the distance deduced from asteroseismological results is consistent with the inverse parallax derived distance from Gaia DR3 within the uncertainty, implying relatively good self-consistency. It is noted that there is a 9\% distance discrepancy. We tend to think that it could be caused by the large light-curve variation. The apparent magnitude in the $V$ band of 16.152 is adopted here. However, considering a 0.2--0.3\,mag variation, $m_{V}$ could be 16.352\,mag, which would lead to a asteroseismological distance of $\sim 70$\,pc, which equals the astrometric distance of 70\,pc from Gaia DR3.

\section*{Acknowledgments}
The authors thank the referee Dr. A.\,H.\,C{\'o}rsico for valuable suggestions and advices, which improved this work a lot. The authors acknowledge the National Natural Science Foundation of China (NSFC) under grants 12203006, 12033003, 12288102, U1938113, 11773035, 12090044, 12090042, and 11633002. This work is also supported by the Ma Huateng Foundation, the Scholar Program of Beijing Academy of Science and Technology (DZ:BS202002), the Tencent Xplorer Prize, and CSST projects: CMS-CSST-2021-B03, the Innovation Project of Beijing Academy of Science and Technology (11000023T000002062763-23CB059 and 11000022T000000443055), and the International Centre of Supernovae, Yunnan Key Laboratory (No. 202302AN36000101).

A.V.F.’s supernova group at UC Berkeley is grateful for financial support from the Christopher R. Redlich Fund, Briggs and Kathleen Wood (T.G.B. is a Wood Specialist in Astronomy), and numerous other donors. A major upgrade of the Kast spectrograph on the Shane 3\,m telescope at Lick Observatory, led by Brad Holden, was made possible through generous gifts from the Heising-Simons Foundation, William and Marina Kast, and the University of California Observatories. Research at Lick Observatory is partially supported by a generous gift from Google.    
 
This work has made use of data from the European Space Agency (ESA) mission Gaia (\url{https://www.cosmos.esa.int/gaia}), processed by the Gaia Data Processing and Analysis Consortium (DPAC;
\url{https://www.cosmos.esa.int/web/gaia/dpac/consortium}). Funding for the DPAC has been provided by national institutions, in particular the institutions participating in the Gaia Multilateral Agreement.
This research has made use of the SIMBAD database, operated at CDS, Strasbourg, France \citep{Wenger2000}.

\section*{DATA AVAILABILITY}
The TESS data used in this article can be accessed via \url{https://archive.stsci.edu/missions-and-data/tess}. The TMTS, SNOVA photometric, and 3\,m spectral data can be obtained by contacting the corresponding authors.

%%%%%%%%%%%%%%%%%%%%%%%%%%%%%%%%%%%%%%%%%%%

\end{document}